\documentclass[prd,twocolumn,amsmath,superscriptaddress,amssymb,floatfix,nofootinbib]{revtex4}
\usepackage{amsfonts,color,xcolor,graphicx,amsfonts,multirow,amsmath}
\usepackage[ colorlinks=true, citecolor=blue, linkcolor=blue, urlcolor=blue]{hyperref}
\usepackage{ulem}

\newcommand{\DS}[1]{/\!\!\!\!#1}
\allowdisplaybreaks

\begin{document}

\title{Longitudinal leading-twist distribution amplitude of the $^1P_1$-state $b_1(1235)$ meson and its implications on $B\to b_1(1235)\ell^+\nu_\ell$ decays}

\author{Long Zeng}
\email{zlong@cqu.edu.cn}

\author{Xing-Gang Wu}
\email{wuxg@cqu.edu.cn}

	\author{Dan-Dan Hu}
\email{hudd@stu.cqu.edu.cn}

\address{Department of Physics, Chongqing Key Laboratory for Strongly Coupled Physics, Chongqing University, Chongqing 401331, P.R. China}

\author{Hai-Bing Fu}
\email{fuhb@gzmu.edu.cn}

\author{Tao Zhong}
\email{zhongtao1219@sina.com}

\address{Department of Physics, Guizhou Minzu University, Guiyang 550025, P.R.China}
\address{Institute of High Energy Physics, Chinese Academy of Sciences, Beijing 100049, P.R.China}

\begin{abstract}

In the paper, we derive the $\xi$ moments $\langle\xi_{2;b_1}^{n;\|}\rangle$ of the longitudinal leading-twist distribution amplitude $\phi_{2;b_1}^{\|}$ for $^1P_1$-state $b_1(1235)$ meson by using the QCD sum rules under the background field theory. Considering the contributions from the vacuum condensates up to dimension six, its first two nonzero $\xi$ moments at the scale 1 GeV are $\langle\xi_{2;b_1}^{1;\|}\rangle= -0.647^{+0.118}_{-0.113}$ and $\langle\xi_{2;b_1}^{3;\|}\rangle = -0.328^{+0.055}_{-0.052}$, respectively. Using those moments, we then fix the Gegenbauer expansion series of $\phi_{2;b_1}^{\|}$ and apply it to calculate $B\to b_1(1235)$ transition form factors (TFFs) that are derived by using the QCD light-cone sum rules. Those TFFs are then extrapolated to the physically allowable $q^2$ range via the simplified series expansion. As for the branching fractions, we obtain ${\cal B}(\bar B^0 \to b_1(1235)^+e^- \bar\nu_e) = 2.179^{+0.553}_{-0.422}\times 10^{-4}$, ${\cal B}(B^0 \to b_1(1235)^-\mu^+\nu_\mu) = 2.166^{+0.544}_{-0.415}\times 10^{-4}$, ${\cal B}(B^+ \to b_1(1235)^0e^+\nu_e) = 2.353^{+0.597}_{-0.456}\times 10^{-4}$, and ${\cal B}(B^+ \to b_1(1235)^0\mu^+\nu_\mu) = 2.339^{+0.587}_{-0.448}\times 10^{-4}$, respectively.

\end{abstract}

\date{\today}

\maketitle

\section{Introduction}

Semileptonic decays of the $B$ meson into light axial-vector mesons constitute a crucial aspect for understanding nonperturbative effects in weak interactions, which have garnered extensive investigation from both theoretical and experimental groups. In quark models, light axial-vector mesons are categorized into two types. The first type includes the $^1P_1$ states, specifically $b_1(1235)$, $h_1(1170)$, $h_1(1380)$, and $K_{1B}$, which collectively form the $J^{PC} = 1^{+-}$ nonets. The second type comprises the $^3P_1$ states, denoted as $a_1(1260)$, $f_1(1285)$, $f_1(1420)$, and $K_{1A}$, and they form the $J^{PC} = 1^{++}$ nonets. Among these axial-vector mesons, only the isospin triplet state mesons $a_1(1260)$ and $b_1(1235)$ exhibit no mixing phenomena, and their internal structures are relatively clear.

Experimentally, the $a_1(1260)$ meson is well established by the BABAR and the Belle Collaborations, and its quark content has been suggested to be a light-quark pair $q\bar q$ with $q=(u,d)$~\cite{Aubert:2004xg, Aubert:2006dd, Aubert:2006gb, Abe:2007jn, Dhar:1983fr, Wingate:1995hy, Wakayama:2019crb}. The $b_1(1235)$-meson, classified as a $^1P_1$ state within the framework of the usual quark model, was first observed by the HBC Collaboration in 1963~\cite{Abolins:1963zz}. Prior to 1994, it was studied extensively in $p\bar p$~\cite{Bizzarri:1969snf, Baltay:1967zza}, $\gamma p$~\cite{OmegaPhoton:1984ols}, and $p\pi$~\cite{Karshon:1974qi, Chaloupka:1974eg, Gessaroli:1977mt, Evangelista:1980xe, OmegaPhoton:1983vde, Collick:1984dkp, Fukui:1990ki, IHEP-IISN-LANL-LAPP-KEK:1992puu, ASTERIX:1993wam} interactions, with a particular emphasis on measuring its mass and width. Reference~\cite{BESIII:2020nme} suggests that the $D$-meson semileptonic decay process $D\to b_1(1235)\ell\nu_\ell$ may present a unique opportunity to study the nature of $b_1(1235)$. In 2020 and 2024, the BESIII Collaboration conducted a series of measurements on $b_1(1235)$ via such kind of decays, {\it e.g.,} $D^{0(+)}\to b_1(1235)^{-(0)}e^+\nu_e$ and $D_s^+\to b_1(1235)^0e^+\nu_e$~\cite{BESIII:2020jan, BESIII:2023clm, BESIII:2024pwp}. As for the $B$-meson decays into $b_1(1235)$, which can be treated via a similar way as the case of the $D$ meson but are characterized by a significantly larger transition momentum $q^2$, they may yield even more intriguing insights into the dynamics of heavy meson decays and the properties of $b_1(1235)$. Even though there are still no data on it, many theoretical studies have been done in the literature.

Theoretically, in studying the semileptonic decay $B\to b_1(1235)\ell\nu_\ell$, the $B\to b_1(1235)$ transition form factors (TFFs) are crucial components for calculating the related observables. At present, the $B\to b_1(1235)$ TFFs have been investigated under various approaches, such as the perturbative QCD (pQCD)~\cite{Li:2009tx}, the QCD light-cone sum rules (LCSRs)~\cite{Yang:2008xw, Sun:2011ssd}, and the covariant light-front quark model (CLFQM)~\cite{Cheng:2003sm, Kang:2018jzg}. Using those TFFs, the branching ratio of $B\to b_1(1235)\ell\nu_\ell$ is at the order of ${\cal O}(10^{-4})$, which could be precisely measured at the high luminosity $B$ factories. Among those approaches, the LCSR offers an effective framework for determining the nonperturbative parameters of hadronic states and it is applicable in both small and intermediate $q^2$ region, where $q^2$ is the transfer momentum between $B$ and $b_1(1235)$, and then a more accurate LCSR prediction is helpful for a better understanding of the TFFs.

Under the LCSR approach, by using the operator product expansion (OPE) near the light cone $x^2 \approx 0$, the nonperturbative hadronic matrix elements can be parametrized as the light-cone distribution amplitudes (LCDAs) of $b_1(1235)$ with different twist structures. Some initial works on the LCDAs of axial-vector mesons have been done in the literature~\cite{Yang:2005gk, Yang:2007zt}. Ref.\cite{Yang:2005gk} studied the leading twist LCDAs of the light $^{1}P_{1}$-state mesons by using the QCD sum rules (QCDSR)~\cite{Shifman:1978bx, Shifman:1978by} to calculate the Gegenbauer moments, which then extended to deal with the twist-3 LCDAs in Ref.\cite{Yang:2007zt} by further using conformal partial wave expansion. In this paper, we will employ the LCSR approach to deal with the $B\to b_1(1235)$ TFFs and will highlight the longitudinal leading-twist LCDA of $b_1(1235)$ by constructing a new correlator with a left-handed chiral current. Within the LCSR approach, the LCDAs reflect important internal structures of a hadron. Similar to the case of usual light mesons, the longitudinal leading-twist LCDA $\phi_{2;b_1}^\|(x)$ of the $b_1(1235)$ meson can be expanded as a Gegenbauer series,
\begin{equation}
\phi_{2;b_1}^\|(x) = 6x\bar x \sum_n a^{n;\|}_{2;b_1} C_n^{3/2}(\xi),  \label{twist2phi}
\end{equation}
where $\bar x = (1-x)$, $\xi=(2x-1)$, and $a^{n;\|}_{2;b_1}$ are Gegenbauer moments. The LCDA and its Gegenbauer moments are scale dependent, whose magnitude at any scale can be derived by using proper evolution equations from their given magnitudes at some initial scale. Throughout the manuscript, if not specially stated, we will omit the scale dependence in those parameters for convenience. The first Gegenbauer moment has been given in Ref.\cite{Yang:2007zt} at the scale $\mu_0 = 1~{\rm GeV}$, {\it e.g.,} $a_{2;b_1}^{1;\|}(\mu_0) = -0.02(2)$. 

In this work, we will adopt the QCD sum rules under background field theory (BFTSR)~\cite{Novikov:1983gd, Hubschmid:1982pa, Govaerts:1983ka, Huang:1986wm, Huang:1989gv} to calculate the $\xi$ moments $\langle\xi_{2;b_1}^{n;\|}\rangle$ and then fix the Gegenbauer moments of the LCDA via their inner relations. Within the framework of BFTSR, the quark and gluon fields are composed by the QCD background fields and their surrounding quantum fluctuations, and the usual vacuum condensates are described by the classical background fields; to take the background field theory as the foundation for the QCDSR approach, it not only presents a distinct physical picture, but also greatly simplifies the calculation due to its capability of adopting different gauges for quantum fluctuations and background fields.

The remaining parts of the paper are organized as follows. In Sec.~\ref{Sec_II}, we present the calculation procedures for the $\xi$ moments of $b_1(1235)$-meson longitudinal leading-twist LCDA, the $B\to b_1(1235)$ TFFs and the branching ratios. Numerical results and discussions are presented in Sec.~\ref{Sec_III}. Section~\ref{Sec_IV} is reserved for a summary.

\section{Calculation Technology}\label{Sec_II}

\subsection{The $\xi$ moments of $b_1(1235)$-meson longitudinal leading-twist LCDA using the BFTSR}

For self-consistency, we first add a simple introduction to BFT which is the basis for BFTSR approach. More detailed introduction of BFTSR can be found in Ref.\cite{Hu:2021zmy}. The main idea of the background field theory is to describe the non-perturbation effect with the classical background field satisfying the equation of motion and to describe the quantum fluctuation, namely the perturbation effect, on this basis within the frame of quantum field theory. More specifically, one can use the following substitution for the usual QCD Lagrangian:
	\begin{align}
		&\mathcal{A}_\mu^a(x) \to \mathcal{A}_\mu^a(x) + \phi_\mu^a(x), \label{Eq:1}\\
		&\psi(x) \to \psi(x) + \eta(x), \label{Eq:2}
	\end{align}
where $\mathcal{A}_\mu^a(x)$ and $\psi(x)$ stand for the gluon and quark background fields, $\phi_\mu^a(x)$ and $\eta(x)$ represent their quantum fluctuations, respectively. The physical observable is gauge independent. And as an advantage of choosing the background field gauge is that it makes the theory be invariant under the background field gauge, making the calculated physical quantity independent of the gauge. With the help of Eq.\eqref{Eq:1} and  Eq.\eqref{Eq:2}, we can obtain the following effective Lagrangian:
	\begin{align}
		\mathcal{L}_{\mathrm{eff}} &= \mathcal{L}_{\mathrm{QCD}}(\mathcal{A}, \psi) + \mathcal{L}(\mathrm{ghosts}) + \bar{\eta}(i\DS{D} - m)\eta + \frac{1}{2}\phi_{\mu}^{a}
		\nonumber\\
		&\quad \times\left\{ g^{\mu \nu}D_{ac}^{2} - \left( 1 - \frac{1}{\alpha} \right)[D^{\mu}D^{\nu}]_{ac} + 2g_sf^{abc}G_{b}^{\mu \nu} \right\}\nonumber\\
		&\quad \times\phi_{\nu}^{c} + g_s(\bar{\psi}\phi^{a}T^{a}\eta + \bar{\eta}\phi^{a}T^{a}\eta) - g_{s}^{2}f^{adf}f_{abc}\mathcal{A}_{d}^{\mu}\phi_{f}^{\nu} \nonumber\\
		&\quad \times \phi_{\mu}^{b}\phi_{\nu}^{c} - g_sf^{abc}(\partial_{\mu}\phi_{\nu}^{a})\phi_{b}^{\mu}\phi_{c}^{\nu} - \frac{1}{4}g_{s}^{2}f^{abc}f_{acf}\phi_{b}^{\mu}\phi_{c}^{\nu}\nonumber\\
		&\quad \times \phi_{\mu}^{d}\phi_{\nu}^{f} + g_s\bar{\eta}\phi^{a}T^{a}\eta.
	\end{align}
where $\mathcal{L}_{\mathrm{QCD}}(\mathcal{A}, \psi)$ with $\mathcal{A}_\mu^a(x)$, $\psi(x)$ has the usual form of QCD Lagrangian and can be minimized to zero when the classical fields $\mathcal{A}_\mu^a(x)$ and $\psi(x)$ are the solutions of the equation of motion, and $\alpha $ is the gauge-fixing parameter. The $\mathcal{L}(\mathrm{ghosts})$ is the contribution of the ghost particle term. Among them, the gluon quantum field satisfies the background field gauge, 	
	\begin{align}
		&D_\mu^{ab}(\mathcal{A})\phi_b^\mu = 0, \\
		&D_\mu^{ab}(\mathcal{A}) = \delta^{ab}\partial_\mu - g_sf^{abc}\mathcal{A}_\mu^c.
	\end{align}
where the color indices $a,b,c = (1,2,...,8)$, $g_s$ is the coupling constant of strong interactions, $f^{abc}$ is the structure constant of the ${\rm SU}_f(3)$ group. 

The basic definition of the longitudinal leading-twist LCDA of the axial-vector meson $b_1(1235)$ is~\cite{Yang:2005gk, Yang:2007zt}
\begin{align}
&\langle b_1(q,\lambda)|\bar q_1(z)\gamma_\mu \gamma_5 q_2(-z)|0\rangle  \nonumber\\
& = i m_{b_1} f_{b_1}^\| \int_0^1 dx e^{i(xz\cdot q - \bar xz\cdot q)} q_\mu \frac{e^{*(\lambda)}\cdot z}{q\cdot z}\phi_{2;b_1}^\|(x). \label{Eq:phi2}
\end{align}
Here, two light quarks $q_1 = q_2$, which are $(u, d)$ quarks for the $b_1(1235)$ meson, respectively. $f_{b_1}^\|$ is the $b_1(1235)$-meson longitudinal decay constant, which can be derived from the transverse one $f^\bot_{b_1}$ via the relations suggested in Ref.\cite{Yang:2007zt}. $q$ and ${e^{*(\lambda )}}$ are the momentum and polarization vector of the $b_1(1235)$ meson, respectively. The polarization vector satisfies the relationship $(e^{*(\lambda)} \cdot z) /({q \cdot z}) \to 1/{m_{b_1}}$~\cite{Ball:2004rg}. By performing the series expansion on both sides of Eq.\eqref{Eq:phi2}, we obtain
\begin{align}
&\langle 0| \bar q_1(0)/\!\!\! z\gamma_5  {(iz\cdot\tensor D)^{n}} q_2(0)|b_1(q,\lambda)\rangle
\nonumber\\
&\hspace{3.5cm} =i{(z\cdot q)}^{n+1} f_{b_1}^\| \langle\xi^{n;\|}_{2;b_1}\rangle,
\end{align}
where the $\xi$ moments are defined as
\begin{equation}
\langle\xi^{n;\|}_{2;b_1}\rangle = {\int_{\rm{0}}^{\rm{1}} {dx(2x - 1)} ^n}\phi _{2;{b_1}}^\parallel (x).
\end{equation}
The covariant derivative satisfies the relation $(iz \cdot \tensor D)^n=(iz\cdot \overrightarrow D - iz\cdot \overleftarrow D)^n$. Because of the conservation of $G$ parity, the $^1P_1$-state axial-vector meson's chiral-even longitudinal leading-twist LCDA $\phi_{2;b_1}^\|(x)$ is antisymmetric under the replacement $x\to 1-x$ in the SU(3) limit, which follows the normalization $\int_0^1 dx \phi_{2;b_1}^\|(x) =0$. Meanwhile, the twist-three LCDA $\psi_{3;b_1}^\|(x)$ is symmetric for the $^1P_1$-state axial-vector meson, {\it e.g.,}
\begin{align}
& \langle b_1(q,\lambda)|\bar q_1(z)\gamma_5 q_2(-z)|0\rangle
\nonumber\\
&\qquad= m_{b_1}^2 f_{b_1}^\bot(\epsilon^{*(\lambda)}\cdot z) \int_0^1 dx e^{i(xz\cdot q - \bar xz\cdot q)} \frac{\psi_{3;b_1}^\|(x)}{2},
\end{align}
which satisfies the normalization $\int_0^1 dx \psi_{3;b_1}^\|(x) =1$~\cite{Yang:2007zt}.

We construct the following correlation function (correlator) to derive the QCD sum rules for $\xi$ moments:
\begin{eqnarray}
\Pi_{2;b_1}^{\|;(n,0)}(z,q^2) &=& i \int d^4x e^{iq\cdot x}\langle 0 |T\{ J_{2;b_1}^{\|;n}(x),J_{3;b_1}^{\|;0 \dagger}(0)\}|0\rangle
\nonumber \\
&=&(z\cdot q)^{n+2} I_{2;b_1}^{\|,(n,0)}(q^2), \label{Eq:Correlator1}
\end{eqnarray}
with $z^2\rightsquigarrow 0$, $J_{2;b_1}^{\|;n}(x)= \bar q_1(x)/\!\!\! z\gamma_5 (iz\cdot\tensor D)^n q_2(x)$, and $J_{3;b_1}^{\|;0 \dagger}(0)= \bar q_2(0)\gamma_5 q_1(0)$. The $^1P_1$-state LCDA $\phi _{2;b_1}^\|(x)$ is defined by the antisymmetric nonlocal axial-vector current, indicating that only odd moments, {\it i.e.,} $n=(1,3,5,\cdots)$, are nonzero.

On the one hand, in the deep Euclidean region $q^2 \ll 0$, one can apply the OPE for the correlator~(\ref{Eq:Correlator1}) using the BFTSR approach. Subsequently, the correlator can be calculated with the help of three key components, {\it e.g.,} the two quark propagators $S_F^d(0,x)$ and $S_F^u(x,0)$, along with the vertex operator $(iz\cdot \tensor{D})^{n}$, whose expressions up to dimension-six condensates can be found in Ref.\cite{Hu:2021zmy}.

On the other hand, the correlation function (correlator) $\Pi_{2;b_1}^{\|;(n,0)}(z,q^2)$ in the physical region, one can insert a complete set of intermediate hadronic states, including the ground state of the $b_1(1235)$ meson into the correlator, {\it i.e.,}
\begin{align}
{\rm Im} I_{2;b_1,{\rm Had}}^{(n,0)}(q^2)&=\pi \delta(q^2-m_{b_1}^2) f_{b_1}^\| f_{b_1}^\bot \langle\xi_{2;b_1}^{n;\|}\rangle \langle\xi_{3;b_1}^{0;\|}\rangle  \nonumber\\
&+\frac{3}{4\pi(n+1)(n+3)}\theta (q^2-s_{0}),  \label{rm}
\end{align}
where $s_{0}$ is the continuum threshold. The first (second) terms on the right-hand side of Eq.\eqref{rm} are contributions of the $b_1(1235)$-meson ground state (continuum states), respectively. $\langle\xi_{3;b_1}^{0;\|}\rangle$ represents the $0_{\rm th}$ $\xi$ moment of two-particle twist-three LCDA $\psi_{3;b_1}^\|(x)$. Since not all the contributions from the higher-order QCD corrections and higher-dimensional operators have been included, the fixed-order prediction of $\langle\xi_{3;b_1}^{0;\|}\rangle$ will be close to but not exactly equal to $1$. So, we will reserve the term $\langle\xi_{3;b_1}^{0;\|}\rangle$ in the hadronic expression \eqref{rm} and its magnitude will be calculated from its own sum rule.

The correlator is analytic in the whole $q^2$ region, so one can bridge the invariant function and the OPE side by using the dispersion relation. Furthermore, the Borel transformation is used to suppress the contributions from the unknown continuum states and high-dimension condensates, {\it e.g.,}
\begin{eqnarray}
\frac{1}{\pi M^2}\int ds e^{-s/M^2}\textrm{Im}I_{b_1,\textrm{Had}}^{(n,0)}(s)=\hat {\cal B}_{M^2} I_{2;b_1,{\rm QCD}}^{(n,0)}(q^2),
\label{Eq:dispersion relation}
\end{eqnarray}
where $M^2$ is the Borel parameter, which comes from the Borel transformation
\begin{equation}
\hat{\cal B}_{M^2}=\underset{\begin{array}{c}
    -q^2,n\rightarrow \infty\\
    -q^2/n=M^2\\
\end{array}}{\lim}\frac{1}{( n-1) !}(-q^2) ^n\left[ -\frac{d}{d( -q^2 )} \right]^n .
\end{equation}

Using the dispersion relation and performing the Borel transformation, we obtain the following QCD sum rules for $\langle\xi_{2;b_1}^{n;||}\rangle \langle\xi_{3;b_1}^{0;||}\rangle$ up to dimension-six condensates,
\begin{widetext}
\begin{eqnarray}
&&\hspace{-0.5cm}\frac{\langle\xi^{n;||}_{2;b_1}\rangle \langle \xi^{0;||}_{3;b_1}\rangle m_{b_1} f_{b_1}^\| f_{b_1}^\bot} { e^{m_{b_1}^2/M^2}} ~=~ - \frac{6 m_q}{8\pi^2(n+2)}~ \left(1-e^{-s_0/M^2}\right) ~+ 4 \langle\bar qq \rangle + \frac{\langle g_s\bar q q \rangle^2 }{81M^4} 4m_q (n+3)  -  \frac{\langle g_s^2\bar qq\rangle^2}{972 \pi^2 M^4} m_q(2+\kappa^2)
\nonumber\\
&& \qquad \times \bigg\{\delta^{0n}\Big[-24\,\bigg(-\ln\frac{M^2}{\mu^2}\bigg)-148\Big]\,+\,\delta^{1n}\bigg[128 \bigg(-\ln\frac{M^2}{\mu^2}\bigg)- 692\bigg] \,+\, \theta(n-1)\bigg[8\,(6n^2+34n)\,\bigg(- \ln\frac{M^2}{\mu^2}\bigg)
\nonumber\\
&& \qquad+4n\tilde\psi(n)-2(6n^2+96n+212)\bigg]+\theta(n-2)\bigg[8(33n^2-17n)
\bigg(-\ln\frac{M^2}{\mu^2}\bigg) - 2(6n^2 +71n)\tilde\psi(n) - \frac{1}{n(n-1)}
\nonumber\\
&& \qquad \times  (231n^4\,+520n^3\,-1101n^2\,+230n)\,\bigg]\, +\, \theta(n-3) \bigg[(74n-144n^2)\, \tilde\psi(n) \,-\, \frac{1}{n-1}\, (169n^3-348n^2+245n
\nonumber\\
&& \qquad +60)\bigg]+4(n+5)\bigg\} \,-  \frac{\langle\alpha_s G^2\rangle}{12\pi M^2} m_q \bigg\{ 12n \bigg( -\ln\frac{M^2}{\mu^2} \bigg) -6(n+2) +\theta(n-1) \bigg[4n\bigg(-\ln\frac{M^2}{\mu^2} \bigg) + 3\tilde\psi(n)-\frac6n\bigg]
\nonumber\\
&&\qquad
+\theta(n-2)\bigg[-(8n+3) \tilde \psi(n) - 2(2n+1) + \frac6n \bigg] \bigg\} -  \frac{ \langle g_s^3 f G^3 \rangle}{96 \pi^2 M^4} m_q  \bigg\{ \delta^{1n} \bigg[-24 \bigg( - \ln \frac{M^2} {\mu^2} \bigg)+84\bigg]+\theta(n-1)
\nonumber\\
&&\qquad
\times\bigg[-4n(3n-5)\,\bigg(-\ln\frac{M^2}{\mu^2}\bigg) \,+\, 2 (2n^2 \,+5n \, -13)\bigg]\,+\,\theta(n-2)\,\bigg[-24n^2\bigg(-\ln\frac{M^2}{\mu^2}\bigg) \,+\, 2n\,(n-4) \, \tilde\psi(n)
\nonumber\\
&&\qquad  + 17n^2 +55n+12\bigg] +\theta(n-3)\bigg[2n(n-4)\tilde\psi(n)+\frac{1}{n-1}(19n^3-32n^2+7n+6)\bigg] \bigg\} -  \frac{\langle{g_s}\bar q \sigma TGq\rangle }{3M^2} 8n,
\label{Eq:SRxi2nxi30}
\end{eqnarray}
\end{widetext}
where $\mu$ is the renormalization scale, $\kappa=\langle\bar s s\rangle / \langle\bar q q\rangle$, and $\tilde\psi(n)=\psi(\frac{n+1}2) - \psi(\frac n2) +(-1)^n\ln4$. The sum rules for the $b_1(1235)$-meson twist-three LCDA's zeroth moment $\langle \xi^{0;||}_{3;b_1}\rangle$ can be derived from the following correlator,
\begin{displaymath}
\Pi^{\|;(0,0)}_{3;b_1} (z,q^2) = i \int d^4x e^{iq\cdot x} \langle 0| T \{ J_{3;b_1}^{\|;0}(x), J_{3;b_1}^{\|;0 \dag}(0)\} |0\rangle.
\end{displaymath}
Following similar QCDSR procedures, we obtain
\begin{align}
    &\frac{\langle \xi^{0;\|}_{3;b_1}\rangle^2 m_{b_1}^2 (f_{b_1}^\bot)^2}{M^2e^{m_{b_1}^2/M^2}}\!=\! \frac{3}{2\pi^2(n+1)}\!\bigg[M^2\!-\!(M^2\!+\!s_0)e^{- \frac{s_0}{M^2}}\!\bigg]
\nonumber\\
    &\hspace{0.8cm}  + \frac{\langle\alpha_s G^2\rangle}{2\pi M^2} +\frac{\langle g_s^2\bar qq\rangle^2 (2+\kappa^2)}{243\pi^2M^4}\bigg[70-32\left(-\ln \frac{M^2}{\mu^2}\right)\bigg]
\nonumber\\
    &\hspace{0.8cm}  - \frac{32\,\langle g_s\bar q q\rangle^2}{27M^4} ~+~ \frac{4\langle g_s\bar q \sigma TGq\rangle}{M^4}\,+\,12m_q~\frac{\langle\bar q q\rangle}{M^2},
\label{Eq:xi30xi30}
\end{align}
where $q=(u,d)$. Using Eqs. \eqref{Eq:SRxi2nxi30} and \eqref{Eq:xi30xi30}, we then obtain
\begin{equation}
	\langle\xi^{n;\|}_{2;b_1}\rangle=\frac{(\langle\xi^{n;\|}_{2;b_1}\rangle \langle\xi^{0;||}_{3;b_1} \rangle)}{\sqrt{\langle\xi^{0;||}_{3;b_1} \rangle^2} }.\label{Eq:xi2n}
\end{equation}
It is found that, by using Eq.~(\ref{Eq:xi2n}), one can eliminate some systematic errors from input parameters and achieve more precise prediction on the $\xi$ moments~\cite{Zhong:2021epq, Zhong:2022ecl}.

\subsection{The $B \to b_1(1235)$ TFFs using the LCSR}

We adopt the following correlator to derive the LCSRs for the $B \to b_1(1235)$ TFFs,
\begin{eqnarray}
\Pi_\mu (p,q) &=& i\int d^4x e^{iq\cdot x}\langle b_1(p,\lambda) |T\{j_\mu(x),j_B^\dag(0)\}|0\rangle
\nonumber \\ [1.5ex]
&=&  -\Pi_1 e_\mu^{*(\lambda )} + \Pi_2(e^{*(\lambda)}\cdot q)(2p+q)_\mu
\nonumber \\ [1.5ex]
&& +  \Pi_3(e^{*(\lambda)}\cdot q){q_\mu } + i{\Pi _V} \epsilon_\mu^{\alpha\beta\gamma} e_\alpha^{*(\lambda )} q_\beta p_\gamma,
\label{cf}
\end{eqnarray}
where $p$ is the $b_1(1235)$-meson momentum, $q=p_B-p_{b_1}$ is the momentum transfer, and ${e^{*(\lambda )}}$ is $b_1(1235)$-meson polarization vector with $\lambda = (\bot,\|)$ being its transverse or longitudinal component, respectively. $j_\mu(x)=\bar q_2(x)\gamma_\mu (1-\gamma_5 )b(x)$ and $j_B^\dag (x) = i\bar b(x)(1 - {\gamma_5 })q_1(x)$.

In the timelike $q^2$ region, the long distance quark-gluon interactions are dominant. To deal with the correlator in the timelike region, one can insert a complete set of the $B$-meson states, which have the same $J^P$-quantum numbers to obtain the hadronic expression. After separating the $B$-meson pole term, we obtain
\begin{align}
&\hspace{-0.25cm}\Pi_\mu^{\rm H}(p,q)  = \frac{{\langle b_1|\bar q_2 \gamma_\mu (1-\gamma_5) b |B \rangle \langle B|\bar b i \gamma_5 q_1|0\rangle }}{m_B^2 - (p+q)^2}
\nonumber\\
&+\sum\limits_{\rm H}\frac{\langle b_1|\bar q_2\gamma_\mu (1-\gamma_5) b|B^{\rm H}\rangle \langle B^{\rm H}|\bar bi(1-\gamma_5)q_1|0\rangle }{m_{B^{\rm H}}^{2} - (p+q)^2},
\end{align}
where $\langle B|\bar b i\gamma_5 q|0\rangle = m_B^2 f_B/m_b$. The $B\to b_1(1235)$ transition matrix elements have the expressions~\cite{Li:2009tx},
\begin{align}
&\langle b_1(p,\lambda)|\bar q_2 \gamma_\mu \gamma_5  b|B(p+q)\rangle \!=\! -\epsilon^{\mu\nu\alpha\beta}e_\nu^{*(\lambda )}q_\alpha p_\beta
\frac{2iA(q^2)}{m_B- m_{b_1}},
\nonumber\\ \\\label{Eq:DefA}
&\langle b_1(p,\lambda)|\bar q_2 \gamma_\mu b|B(p+q)\rangle = - e_\mu ^{*(\lambda )}(m_B- m_{b_1})V_1(q^2)
\nonumber\\
&\hspace{1.5cm} + (2p+q)_\mu \frac{e^{*(\lambda)} \cdot q}{m_B- m_{b_1}} V_2(q^2)
\nonumber \\[1.5ex]
&\hspace{1.5cm} + q_\mu (e^{*(\lambda)}\cdot q) \frac{2m_{b_1}}{q^2}  [V_3(q^2)- V_0(q^2)].
\end{align}
There is one linear relationship among the TFFs~\cite{Wang:2008bw, Verma:2011yw},
\begin{align}
&&\hspace{-1cm} V_3(q^2)= \frac{m_B- m_{b_1}}{2m_{b_1}} V_1(q^2) - \frac{m_B+ m_{b_1}}{2m_{b_1}} V_2(q^2).
\end{align}
Following standard procedures of QCD sum rules, one can represent the contributions of higher resonances and continuum states by dispersion integrations so as to derive the expressions for the hadronic invariant amplitudes $\Pi_{i}^{\rm H}[q^2,(p+q)^2]$ with $i = (1,2,3,V)$ defined in Eq.(\ref{cf}). The continuum threshold parameter $s_0$ is usually set as the value close to the squared mass of the first excited state of the $B$ meson. Meanwhile, the conventional quark-hadron duality ansatz, $\rho_{i}^\textrm{Had} = \rho_{i}^\text{QCD}\theta(s-s_0)$, can be used to calculate the hadron spectrum density $\rho_{i}^\textrm{had}$.

On the other hand, in spacelike region, one can calculate the correlator via the QCD theory. In this region, the correlator can be treated by the OPE with the coefficients being pQCD calculable. The $b$-quark propagator that shall be used in the calculation can be found in Ref.~\cite{Hu:2021zmy}. After applying the OPE and using the expressions for the transition matrix elements, one can arrange the resultant expressions by twist-two, -three, and -four LCDAs~\cite{Yang:2007zt, Momeni:2016kjz}. After matching the correlator with the dispersion relation and applying the Borel transformation to suppress the less known continuum contributions, the resultant TFFs under the LCSR approach are
\begin{widetext}
\begin{eqnarray}
&&\hspace{-1cm} V_1(q^2) = \frac{2m_b^2m_{b_1}f_{b_1}^\|}{m_B^2f_B(m_B - m_{b_1})} ~\int_0^1\frac{du}{u} ~e^{(m_B^2-s(u))/{\cal M}^2} ~\bigg[\Theta(c(u,s_0))\phi_{3;b_1}^\bot (u) - \widetilde\Theta(c(u,s_0))\frac{m_{b_1}^2}{u{\cal M}^2}\Psi_{4;b_1}^\|(u)\bigg] + \frac{2m_b^2m_{b_1}^2}{m_B^2 f_B}
\nonumber \\
&&\hspace{-1cm}
\qquad\quad\times \frac{(f_{3;b_1}^V - f_{3;b_1}^A)}{(m_B- m_{b_1})}\int {{\rm{\cal D}}\alpha_i } \int_0^1 dv e^{(m_B^2-s(X))/{\cal M}^2}  \frac1{X^2{\cal M}^2} \Theta(c(X,s_0)) \left[\tilde \Phi_{3;b_1}^\|(\alpha_i)-\Phi_{3;b_1}^\|(\alpha_i)\right],
\label{Eq:V1q2}
\\
&&\hspace{-1cm}
V_2(q^2) = \frac{2\,m_b^2\,m_{b_1}\,f_{b_1}^\|\,(m_B- m_{b_1})}{m_B^2f_B}\int_0^1 \hspace{0.1cm} \frac{du}{u}e^{(m_B^2-s(u))/{\cal M}^2} ~\bigg[ \frac1{u {\cal M}^2} \widetilde\Theta(c(u,s_0)) ~\Phi_{2;b_1}^\|(u) + \frac{m_{b_1}^2}{u{\cal M}^4} \widetilde{\widetilde\Theta}(c(u,s_0))
\Psi_{4;b_1}^\|(u)\hspace{-0.01cm} \bigg]
\nonumber \\
&&\hspace{-1cm} \qquad\quad -\frac{m_b^2m_{b_1}^2(f_{3;b_1}^V - f_{3;b_1}^A)(m_B- m_{b_1})}{m_B^2 f_B}\int ~{\cal D}\alpha_i \!\int_0^1 dv e^{(m_B^2-s(X))/{\cal M}^2} \frac1{X^3 {\cal M}^4} \Theta(c(X,s_0)) \bigg[\tilde \Phi_{3;b_1}^\| (\alpha_i )- \Phi _{3;b_1}^\| (\alpha_i )\bigg],
\label{Eq:V2q2}
\\
&&\hspace{-1cm}  V_0(q^2)=V_3(q^2)\!-\!\bigg\{ -\frac{q^2m_b^2f_{b_1}^\|}{m_B^2 f_B}\int_0^1 \frac{du}{u} e^{(m_B^2- s(u)) / {\cal M}^2} \bigg[\frac1{u{\cal M}^2} \widetilde\Theta(c(u,s_0)) \Phi_{2;b_1}^\|(u) - \frac{m_{b_1}^2(2-u)}{u^2 {\cal M}^4}\widetilde{\widetilde\Theta}(c(u,s_0))\hspace{-0.02cm}\Psi_{4;b_1}^\|(u) \bigg]
\nonumber \\
&&\hspace{-1cm} \qquad\quad +\hspace{0.2cm}\frac{q^2~m_b^2\,m_{b_1}~(f_{3;b_1}^V \,-\, f_{3;b_1}^A)}{m_B^2{f_B}}\int ~{\cal D}\alpha_i ~\int_0^1 dv ~ e^{(m_B^2-s(X))/{\cal M}^2}\,\frac1{X^3{\cal M}^4} \Theta(c(X,s_0))~\bigg[~\tilde\Phi_{3;b_1}^\| (\alpha_i)- \Phi_{3;b_1}^\|(\alpha_i )\hspace{0.1cm}\bigg]\bigg\},
\label{Eq:V3q2}
\\
&&\hspace{-1cm} A(q^2)=\frac{m_b^2m_{b_1}f_{b_1}^\|(m_B- m_{b_1})}{2m_B^2 f_B}\int_0^1 du e^{(m_B^2-s(u))/{\cal M}^2} \frac1{u^2{\cal M}^2}\widetilde\Theta(c(u,s_0))\psi_{3;b_1}^\bot (u),
\label{Eq:Aq2}
\end{eqnarray}
\end{widetext}
where the $q^2$-dependent factors $s(u)$ and $s(X)$ are defined as $s(\zeta)=[m_b^2 - (1 - \zeta)({q^2} - \zeta m_{b_1}^2)]/\zeta$ with $\zeta = (u,X)$, where $X=\alpha _1 -\alpha _2 + v \alpha _3$, and $\alpha_1, \alpha _2$, and $\alpha_3$ are momentum fractions carried by ${\bar q}_1$, $q_2$ quarks and gluons in the $b_1(1235)$ meson~\cite{Yang:2007zt}, respectively. $\Theta(c(u,s_0))$ is the usual step function, and $\widetilde\Theta (c(u,s_0))$ and $\widetilde{\widetilde\Theta}(c(u,s_0))$ are defined as
\begin{align}
&\int_0^1 \frac{du}{u^2 {\cal M}^2} e^{-s(u)/{\cal M}^2}\widetilde\Theta(c(u,s_0))f(u)
\nonumber\\
&\qquad = \int_{u_0}^1\frac{du}{u^2 {\cal M}^2} e^{-s(u)/{\cal M}^2}f(u) + \delta(c(u_0,s_0)),
\label{Theta1}\\
&\int_0^1 \frac{du}{2u^3 {\cal M}^4} e^{-s(u)/{\cal M}^2}\widetilde{\widetilde\Theta}(c(u,s_0))f(u)
\nonumber\\
&\qquad = \int_{u_0}^1 \frac{du}{2u^3 {\cal M}^4} e^{-s(u)/{\cal M}^2}f(u)+\Delta(c(u_0,s_0)), \label{Theta2}
\end{align}
where $c(u,s_0)=u s_0 - m_b^2 + \bar u q^2 - u \bar u m_{b_1}^2$ and
\begin{align}
&\delta(c(u,s_0))= e^{-s_0/{\cal M}^2}\frac{f(u_0)}{m_b^2 + u_0^2 m_{b_1}^2 - q^2}, \\
&\Delta(c(u,s_0))= e^{-s_0/{\cal M}^2}\bigg[\frac{1}{2 u_0 {\cal M}^2}\frac{f(u_0)} {m_b^2 + u_0^2 m_{b_1}^2 - q^2} \nonumber\\
&\left. -\frac{u_0^2}{2(m_b^2 + u_0^2 m_{b_1}^2 - q^2)} \frac{d}{du}\left( \frac{f(u)}{u(m_b^2 + u^2m_{b_1}^2 - q^2)} \right) \right|_{u = {u_0}}\bigg].
\end{align}
Here $u_0$ is the solution of $c(u_0,s_0) = 0$ with $0\leq u_0 \leq 1$. The simplified LCDAs are defined as
\begin{align}
&\Phi_{2;b_1}^\| (u) = \int_0^u dv\phi _{2;b_1}^\| (v),
\\
&\Phi_{3;b_1}^\bot (u) = \int_0^u {dv} \phi_{3;b_1}^\bot (v),
\\
&\Psi_{4;b_1}^\|(u) = \int_0^u {dv} \int_0^v {dw} \psi _{4;b_1}^\|(w).
\end{align}
The coupling constants $f_{3;b_1}^V$ and $f_{3;b_1}^A$ for the $b_1(1235)$ meson are defined via the following matrix elements:
\begin{align}
&\langle b_1(q,\lambda)|J_{3,\mu}^{3,A}(0)|0 \rangle = -f_{3;b_1}^A(z\cdot q)^3e^{(\lambda)}_{\bot,\mu}+{\cal O}(z_\mu),\nonumber \\
&\langle b_1(q,\lambda)|J_{3,\mu}^{1,V}(0)|0 \rangle = -if_{3;b_1}^V(z\cdot q)^2e^{(\lambda)}_{\bot,\mu}+{\cal O}(z_\mu),
\end{align}
in which the two interpolating currents $J_{3,\mu}^{3,A}(0)=z^{\alpha}z^{\beta}\bar q_2(0)\gamma_{\alpha}\gamma_5[G_{\beta\mu}(0)i(z\cdot \overrightarrow D)- i(z\cdot \overleftarrow D) G_{\beta\mu}(0)]q_1(0)$ and $J_{3,\mu}^{1,V}(0)=z^{\alpha}z^{\beta}\bar q_2(0)\gamma_{\alpha}g_s{\tilde G_{\beta \mu }}(0)q_1(0)$. ${\cal O}(z_\mu)$ represent the twist-four and higher-twist corrections~\cite{Yang:2007zt}.

In the standard model, the effective Hamiltonian for semileptonic decays $B\to b_1\ell\bar\nu_\ell$ can be written as
\begin{align}
{\cal H}_{\rm eff} = \frac{G_F}{\sqrt{2}}V_{ub}\bar u\gamma_\mu (1-\gamma_5)b\bar\ell \gamma^\mu (1-\gamma_5)\nu_\ell,
\end{align}
where $V_{ub}$ is the Cabibbo-Kobayashi-Maskawa (CKM) matrix element and $G_F$ is the Fermi constant.

Then, the longitudinal and transverse differential decay widths for semileptonic decay $B\to b_1(1235)\ell^+\nu_\ell$ can be expressed as
\begin{eqnarray}
&&\hspace{-1cm}~~\frac{d\Gamma_{\rm L}(B\to b_1(1235)\ell^+\nu_\ell)}{dq^2}
= \bigg(\frac{q^2 - m_\ell^2}{q^2}\bigg)^2\frac{\sqrt\lambda G_F^2|V_{ub}|^2}{384\pi ^3m_B^3}
\nonumber \\
&&\hspace{-1cm}~~\quad\times\frac{1}{q^2}\bigg[3\,m_\ell^2\,\lambda\,V_0^2(q^2) + \frac{m_\ell^2 + 2q^2}{2m_{b_1}}\,\Big|(m_B^2 - m_{b_1}^2- q^2)
\nonumber \\
&&\hspace{-1cm}~~\quad\times  (m_B- m_{b_1})V_1(q^2)- \frac{\lambda}{m_B- m_{b_1}}{V_2}(q^2)\Big|^2\bigg],
\label{Eq:dgamma1}
\\
&&\hspace{-1cm}~~\frac{d\Gamma_\pm(B\to b_1(1235)\ell^+\nu_\ell)}{dq^2}  = \bigg(\frac{q^2 - m_\ell^2}{q^2}\bigg)^2\frac{\lambda^{3/2}  G_F^2|{V_{ub}}|^2}{384\pi^3m_B^3}
\nonumber \\
&&\hspace{-1cm}~~\quad\times\!(m_\ell^2 + 2q^2)\bigg|\frac{A(q^2)}{m_B- m_{b_1}}\!\mp\!\frac{(m_B- m_{b_1}){V_1}(q^2)}{\sqrt \lambda }\bigg|^2,
\label{Eq:dgamma2}
\end{eqnarray}
where $\lambda  = (m_B^4 + m_{b_1}^4 + q^4 - 2m_{b_1}^2m_B^2 -2q^2m_{B}^2- 2q^2m_{b_1}^2)$, $q^2$ is the squared momentum of the lepton pair, and $m_\ell$ represents the mass of charged lepton. The total differential decay width is $d\Gamma_{\rm L}+d\Gamma_{\rm T}$, in which $d\Gamma_{\rm L}$ and $d\Gamma_{\rm T}=d\Gamma_++d\Gamma_-$ correspond to longitudinal and transverse parts, respectively.

\section{Numerical results and discussions}\label{Sec_III}

To do the numerical calculation, the input parameters are taken as follows. According to the Particle Data Group (PDG)~\cite{Zyla:2020zbs}, we take the $b$-quark mass ${\bar m_b}({\bar m_b}) = 4.18(2)~ {\rm GeV}$ and the masses of the light quarks ${\bar m_u}(2~{\rm GeV}) = 2.16_{-0.26}^{+0.49}$ and ${\bar m_d}(2~{\rm GeV}) = 4.67_{-0.17}^{+0.48}~{\rm MeV}$. The $B$- and $b_1(1235)$-meson masses are $m_{B^0}=5.27972, m_{B^+}=5.27942$ and $m_{b_1} \approx 1.2295~{\rm GeV}$, accordingly. The $B$- and $b_1(1235)$-meson decay constants $f_{B^0}=f_{B^+}=0.214(12)$ and $f_{b_1}^\|=0.258(25) ~{\rm GeV}$~\cite{Mutuk:2018lki}. The typical scale of the $B\to b_1(1235)$ decay processes is $\mu_k = (m_B^2 - m_{b_1}^2)^{1/2}$. Furthermore, the nonperturbative vacuum condensates are the significant parameters to the sum rule, and we take~\cite{Colangelo:2000dp, Narison:2014wqa, Zhong:2021epq}
\begin{eqnarray}
\langle \bar qq\rangle |_{2~{\rm GeV}} &=& (-289.14^{+9.34}_{-4.47})^3~{\rm MeV}^3, \nonumber\\
\langle g_s\bar q\sigma TGq\rangle |_{2~{\rm GeV}} &=&(-1.934_{-0.103}^{+0.188})\times  10^{-2}~{\rm GeV}^5, \nonumber\\
\langle g_s\bar qq\rangle^2 |_{2~{\rm GeV}} &=& (2.082_{-0.697}^{+0.734})\times  10^{-3} ~{\rm GeV}^6, \nonumber\\
\langle g_s^2\bar qq\rangle^2 |_{1~{\rm GeV}} &=& (7.420_{-2.483}^{+2.614})\times  10^{-3}~{\rm GeV}^6, \nonumber\\
\langle \alpha_s G^2 \rangle |_{1~{\rm GeV}} &=& 0.038(11)~{\rm GeV}^4, \nonumber\\
\langle g_s^3fG^3\rangle |_{1~{\rm GeV}} &\approx& 0.045 ~{\rm GeV}^6, \nonumber\\
\kappa &=&  0.74(3).
\end{eqnarray}
The values assigned to gluon condensates are predominantly utilized within the framework of QCD sum rules. The double-gluon condensate $\langle \alpha_s G^2 \rangle$ is deduced from the sum rules pertaining to charmonium, whereas the triple-gluon condensate $\langle g_s^3fG^3\rangle$ finds its basis in the instanton model. A comprehensive discussion on this point can be found in Refs.\cite{Colangelo:2000dp, Shifman:1978bx, Shifman:1978by}. The double-quark condensate $\langle \bar qq\rangle$ and the quark-gluon mixed condensate $\langle g_s\bar q\sigma TGq\rangle$ have been updated in Ref.\cite{Zhong:2021epq} based on the Gell-Mann-Oakes-Renner relation, and the relationship $\langle g_s\bar q\sigma TGq\rangle = m_0^2 \langle q\bar q\rangle$ with $m_0^2=0.80(2)~{\rm GeV}^2$~\cite{Narison:2014ska}. Utilizing the relation $\rho\alpha_s\langle\bar qq\rangle^2 = (5.8\pm1.8)\times10^{-4}~{\rm GeV}^6$ with $\rho \sim (3,4)$~\cite{Narison:2014ska}, one obtains the four-quark condensate $\langle g_s\bar qq\rangle^2$. Similarly, the value of $\langle g_s^2\bar qq\rangle^2$ can be determined by amalgamating this with the updated value of $\langle \bar qq\rangle$. All those scale-dependent parameters, encompassing quark masses and vacuum condensates, should be evolved from an initial scale to the required scale such as $\mu_k$ by employing the renormalization group equations (RGEs) given in Refs.\cite{Yang:1993bp, Hwang:1994vp, Lu:2006fr, Zhang:2021wnv}.

\begin{table}[htb]
\begin{center}
\caption{The first two nonzero longitudinal leading-twist LCDA moments of the $b_1(1235)$ meson at the scale $\mu=\sqrt{M^2}$ within the allowable Borel windows.}
\label{tab:xin}
\begin{tabular}{c c c c c}
\hline
~~$n$~~ &  ~~~~~$M^2$~~~~~              & ~~~~~$\langle\xi_{2;b_1}^{n;\|}\rangle$~~~~~                \\  \hline
1     & $[2.017,3.017]$ & $[-0.562,-0.434]$\\
3     & $[2.272,3.272]$ & $[-0.228,-0.259]$\\
\hline
\end{tabular}
\end{center}
\end{table}

Two pivotal parameters for the QCD BFTSR of $\xi$ moments are continuum threshold $s_0$ and the Borel parameter $M^2$. To ascertain their optimal values and bolster the credibility of sum rule predictions, it is imperative to ensure that the contributions from both the continuum states and the dimension-six condensates remain sufficiently small. The continuum threshold can be established as $s_0=6.06$ GeV$^2$ by confirming the existence of an appropriate Borel window for normalizing $\langle \xi_{3;b_1}^{n;||}\rangle$ in Eq.(\ref{Eq:xi30xi30}). We present the determined Borel windows together with the corresponding $\xi$ moments $\langle\xi_{2;b_1}^{n;\|}\rangle$ at the scale $\mu={\sqrt {M^2}}$ in Table~\ref{tab:xin}. Here, to fix the Borel windows, we have set the continuum contributions to be no more than $(30\%, 35\%)$ for $n=(1,3)$, respectively, which give the upper limits of Borel windows.

If the QCD sum rules (\ref{Eq:SRxi2nxi30}) of $\xi$ moments encompass the contributions from all the dimensional condensates, they will be surely independent of any choices of the Borel parameter $M^2$, indicating the flatness of the $\xi$ moments versus $M^2$. However, for a fixed-order OPE expansion, these $\xi$ moments will vary with $M^2$ within its permissible choices. The extent of this variation heavily depends upon the convergence of the OPE expansion with respect to various powers of $1/M^2$. For the present derived LCSRs, which take the condensates up to dimension-six into consideration, we conservatively stipulate that the fluctuations of $\langle\xi_{2;b_1}^{n;\|}\rangle$ within the Borel window to be less than $10\%$ so as to get the lower limit of the Borel window. A quite small contribution from the dimension-six condensates within this Borel window also ensures the reasonableness of this choice, due to the fact that the coefficient of higher-dimensional condensate is generally suppressed by the proper power of $1/M^2$.

\begin{table}[htb]
\begin{center}
\caption{The first two nonzero leading-twist LCDA Gegenbauer moments of the $b_1(1235)$ meson at the scale $1$ GeV.}
\label{tab:an}
\begin{tabular}{lcc}
\hline
 & $a_{2;b_1}^{1;\|}$ & $a_{2;b_1}^{3;\|}$ \\  \hline
This work                   & $-1.078^{+0.197}_{-0.188}$ & $-0.264^{+0.024}_{-0.018}$\\
QCDSR~\cite{Yang:2007zt}    & $-1.95\pm 0.35$  & ... \\
\hline
\end{tabular}
\end{center}
\end{table}

By taking into account all the mentioned error sources and adopting the RGE, our predictions for the first two $\xi$ moments $\langle\xi_{2;b_1}^{1;\|}\rangle$ and $\langle\xi_{2;b_1}^{3;\|}\rangle$ at $1$ GeV are
\begin{align}
&\langle\xi_{2;b_1}^{1;\|}\rangle|_{1 {\rm GeV}} = -0.647^{+0.118}_{-0.113}, \nonumber\\
&\langle\xi_{2;b_1}^{3;\|}\rangle|_{1 {\rm GeV}} = -0.328^{+0.055}_{-0.052},
\label{Eq:xi}
\end{align}
where the errors are squared averages of the errors from all the mentioned input parameters. Using the relationship between the $\xi^n$ moments and the Gegenbauer moments, one can get the first two nonzero Gegenbauer moments $a_{2;b_1}^{1;\|}$ and $a_{2;b_1}^{3;\|}$ at the same scale, which are listed in Table~\ref{tab:an}.

\begin{figure}[htb]
\begin{center}
\includegraphics[width=0.40\textwidth]{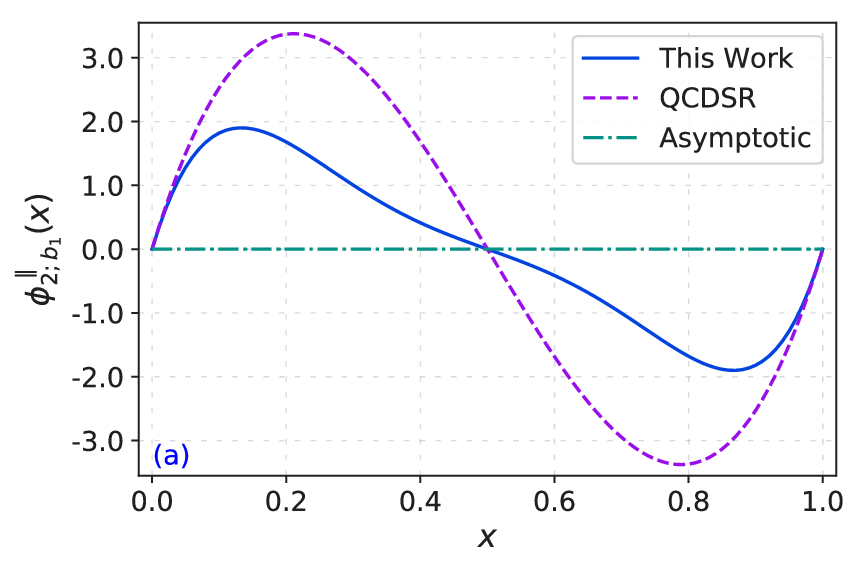}
\end{center}
\caption{The longitudinal leading-twist LCDA $b_1(1235)$ meson $\phi_{2;b_1}^\|(x)$ at the scale $\mu=1$ GeV. The dotted line is the previous QCDSR prediction~\cite{Yang:2007zt}, and the dash-dotted line is asymptotic behavior for $\mu\to\infty$.}
\label{fig:DA1}
\end{figure}

\begin{figure}[htb]
\begin{center}
\includegraphics[width=0.40\textwidth]{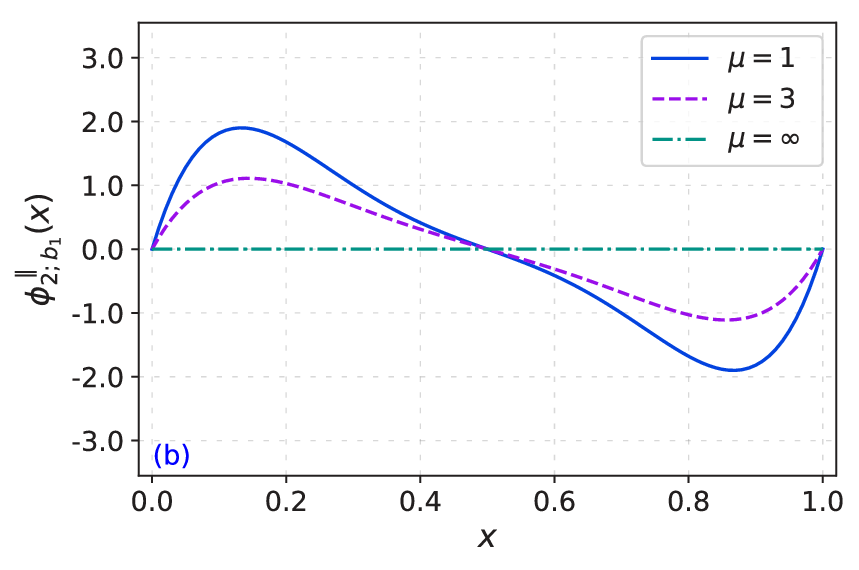}
\end{center}
\caption{The scale-running behavior of the longitudinal leading-twist LCDA $b_1(1235)$ meson $\phi_{2;b_1}^\|(x)$ at $1$ and $3$ GeV. The asymptotic curve for infinity scale ($\infty$) is also presented.}
\label{fig:DA2}
\end{figure}

\begin{figure}[htb]
\begin{center}
\includegraphics[width=0.40\textwidth]{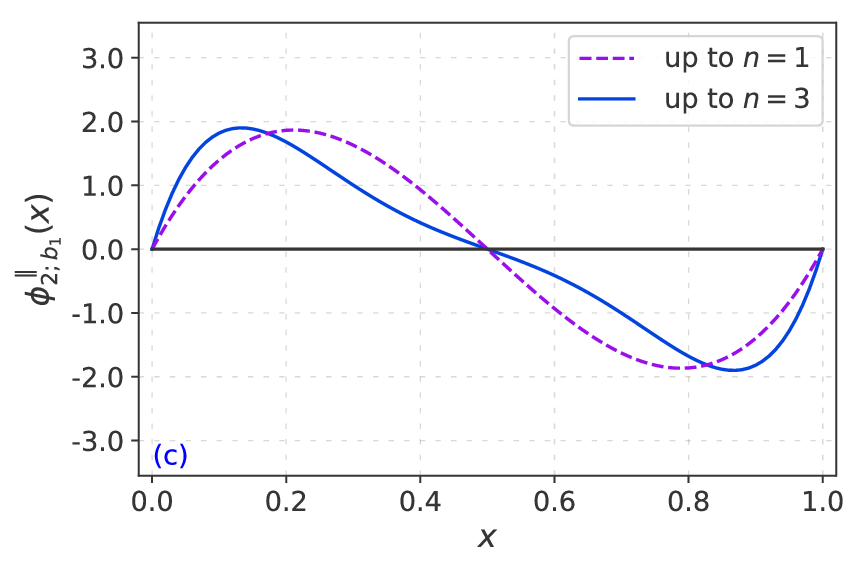}
\end{center}
\caption{The longitudinal leading-twist LCDA $b_1(1235)$-meson $\phi_{2;b_1}^\|(x)$ at $1$ GeV, whose Gegenbauer moments are kept up to $n=1$ and $n=3$, respectively. }
\label{fig:DA3}
\end{figure}

The $B\to b_1(1235)$ TFFs are key elements for investigating the $B$-meson semileptonic decay. To derive the TFFs \eqref{Eq:V1q2}--\eqref{Eq:Aq2}, we need to know explicit forms of the required LCDAs. Regarding the twist-two LCDA $\phi _{2;{b_1} }^\|$, we adopt its truncated Gegenbauer expansion $\phi _{2;{b_1} }^\| (x) = 6x(1 - x)( {a_1}C_1^{3/2}(2x - 1) + {a_3}C_3^{3/2}(2x - 1))$, which is presented in Fig. \ref{fig:DA1}. As a comparison, we also present the QCDSR prediction~\cite{Yang:2007zt} and the asymmetry behavior $\phi_{2;b_1}^{\|}(x)|_{\mu\to\infty} = 6x\bar x$ in Fig. \ref{fig:DA1}. Our prediction prefers an antisymmetric behavior, which is the same as previous QCDSR prediction. Figure \ref{fig:DA2} shows the scale-running behavior of $\phi_{2;b_1}^\|(x)$ at $1$ and $3$ GeV, showing how its shape varies with the increment of scale. Figure \ref{fig:DA3} shows how the LCDA $\phi_{2;b_1}^\|(x)$ at $1$ GeV changes when more Gegenbauer terms are included, {\it e.g.,} the Gegenbauer moments are kept up to $n=1$ and $n=3$, respectively. As for the twist-three LCDAs $\phi_{3;b_1}^\bot(x)$ and $\psi_{3;b_1}^\bot(x)$, they can be decomposed into several parts~\cite{Ball:1998sk},
\begin{align}
&\phi_{3;b_1}^\bot(x) = \phi_{3;b_1}^{\bot WW}(x) + \phi_{3;b_1}^{\bot g}(x) + \phi_{3;b_1}^{\bot m}(x), \\
&\psi_{3;b_1}^\bot(x) = \psi_{3;b_1}^{\bot WW}(x) + \psi_{3;b_1}^{\bot g}(x) + \psi_{3;b_1}^{\bot m}(x).
\end{align}
Here $\phi_{3;b_1}^{\bot WW}(x)$ and $\psi_{3;b_1}^{\bot WW}(x)$ denote the contributions from the leading-twist longitudinal LCDA under the Wandzura-Wilczek approximation, {\it e.g.,}
\begin{eqnarray}
&&\hspace{-1cm} \phi_{3;b_1}^{\bot WW}(x) = \frac12\bigg[\int_0^x \frac{dv}{\bar v}\phi_{2;b_1}^\|(v) + \int_x^1 \frac{dv}v \phi_{2;b_1}^\|(v)\bigg],
\\
&&\hspace{-1cm} \psi_{3;b_1}^{\bot WW}(x) = 2\bar x\int_0^x \frac{dv}{\bar v}\phi_{2;b_1}^\|(v) + 2x\int_x^1 \frac{dv}v \phi_{2;b_1}^\| (v).
\end{eqnarray}
$\phi_{3;b_1}^{\bot g}(x)$ and $\psi_{3;b_1}^{\bot g}(x)$ are contributions from the three-particle distribution amplitudes, which can be safely neglected due to their small contributions. $\phi_{3;b_1}^{\bot m}(x)$ and $\psi_{3;b_1}^{\bot m}(x)$ are related to the coefficient $\tilde\delta_\pm$, which tend to be zero in the $q\bar q$ with $q =(u, d)$ meson components.\footnote{The detailed analysis can be found in Section 4 of Ref.\cite{Ball:1998sk}}

Moreover, $\Phi_{3;b_1}^\|(\alpha_i )$ and $\tilde\Phi_{3;b_1}^\| (\alpha_i)$ represent twist-three LCDAs of the three-particle part~\cite{Ball:1998fj, Cheng:2007mx}. To do the numerical analysis, we adopt~\cite{Ball:1998fj}
\begin{eqnarray}
&&\hspace{-1cm} \Phi_{3;b_1}^\|(\alpha_i )=360\alpha_1\alpha_2\alpha_3^2\bigg[1+\frac{1}2(7\alpha_3-3)\omega_{b_1}^A\bigg],
\\
&&\hspace{-1cm} \tilde\Phi_{3;b_1}^\| (\alpha_i)=5040(\alpha_1-\alpha_2)\alpha_1\alpha_2\alpha_3^2,
\\
&&\hspace{-1cm} \psi_{4;b_1}^\|(x) = 6x\bar x + (1-3\xi^2)\bigg[\frac17 a_{1;b_1}^{2;\|}  - \frac{20}3 \frac{f_{3;b_1}^A}{f_{b_1}^\| m_{b_1}}\bigg].
\end{eqnarray}
where $a_{1;b_1}^{2;\|}=-0.646$ at the scale $\mu$=3 GeV is the result obtained by using the above BFTSR, the coupling constants $f_{3;b_1}^A=-0.0036$ and $f_{3;b_1}^V=0.0030~{\rm GeV^2}$~\cite{Yang:2008xw}, and the coefficient $\omega_{b_1}^A=-1.5 $~\cite{Yang:2008xw}.

In order to evolve the hadronic parameters of $b_1(1235)$-meson twist-two, -three, and -four LCDAs from the initial scale such as $\mu_0$ to the wanted scale $\mu_k$, one can use the RGE with the form $ c_i(\mu_k) = L^{\gamma_{c_i}/\beta _0} c_i(\mu_0)$, where $L = \alpha_s(\mu_k)/\alpha_s(\mu_0)$, $\beta_0 = 11 - 2/3 n_f$. The one-loop anomalous dimensions $\gamma_{c_i}$ satisfy the following equation~\cite{Fu:2020uzy}.
\begin{eqnarray}
\gamma_{c_i} = C_F\bigg(1-\frac2{(n+1)(n+2)}-\sum\limits_{m=2}^{n+1} \frac1m \bigg).
\end{eqnarray}

\begin{table*}[t]
\begin{center}
\caption{The $B\to b_1(1235)$ TFFs at the large recoil point, {\it i.e.,} $A(0), V_1(0), V_2(0)$, and $V_0(0)$. Predictions under various approaches are listed as a comparison.} \label{Tab:TFFs0}
\begin{tabular}{lllll}
\hline
References~~~~~~~~~~~~~~~~~~~~~& $A(0)$ ~~~~~~~~~~~~~~~~~~~~~&$V_1(0)$ ~~~~~~~~~~~~~~~~~~~~~& $V_2(0)$~~~~~~~~~~~~~~~~~~~~~ &  $V_0(0) $   \\   \hline
This~work   & $-0.189^{+0.0549}_{-0.0.068}$   & $-0.096_{-0.017}^{+0.014}$    &$+0.166_{ - 0.023}^{+0.024}$    & $-0.597_{-0.070}^{+0.065}$  \\
LCSR-I~\cite{Yang:2008xw}   & $-0.25^{+0.05}_{-0.05}$   & $-0.20_{-0.04}^{+0.04}$    &$-0.09_{ - 0.02}^{+0.02}$    & $-0.39_{-0.07}^{+0.07}$    \\
LCSR-II~\cite{Sun:2011ssd}   & $-0.17^{+0.06+0.04}_{-0.06-0.04}$   & $-0.29\pm0.09\pm0.06$    &$-0.17_{ - 0.06-0.04}^{+0.06+0.04}$    & $-0.05_{-0.05-0.05}^{+0.05+0.05}$    \\
pQCD~\cite{Li:2009tx}   & $+0.19^{+0.04+0.01+0.03}_{-0.04-0.01-0.03}$   & $+0.33_{-0.06-0.02-0.05}^{+0.07+0.01+0.05}$    &$+0.03_{-0.01-0.00-0.02}^{+0.01+0.00+0.02}$    & $+0.45_{-0.09-0.01-0.04}^{+0.10+0.01+0.04}$     \\
CLFQM~\cite{Cheng:2003sm}   & $+0.10$   & $+0.18$    &$-0.03$    & $+0.39$   \\
\hline
\end{tabular}
\end{center}
\end{table*}

\begin{table}[htb]
\caption{The $B\to b_1(1235)$ TFFs at the large recoil region $q^2=0$, in which the twist-two, -three, and -four LCDA contributions are presented separately.} \label{Tab:twist}
\begin{tabular}{lllll}
\hline
~~~~~~~~~~~~~~~     & $A(0)$~~~~~~~     & $V_1(0)$~~~~~~~    & $V_2(0)$ ~~~~ & $V_0(0)$ ~~~~~ \\   \hline
$\Phi_{2;b_1}^\|(u)$    & $...$           & $...$          & $+0.063$     & $-0.167$    \\
$\phi_{3;b_1}^\bot(u)$  & $...$           & $-0.198$     & $+0.112$           & $-0.622$  \\
$\psi_{3;b_1}^\bot(u)$  & $-0.189$      & $...$          & $...$               & $...$   \\
$\Psi_{4;b_1}^\|(u)$    & $...$           & $+0.102$     & $-0.009$      & $+0.192$   \\
Total                   & $-0.189$      & $-0.096$     & $+0.166$          & $-0.597$ \\
\hline
\end{tabular}
\end{table}

We are ready to derive the TFFs \eqref{Eq:V1q2}--\eqref{Eq:Aq2}, and for the purpose, we need to fix the continuum threshold $s_0$ and Borel parameter ${\cal M}^2$. Normally, the continuum threshold $s_0$ should be taken near the squared mass of the $B$ meson's first excited state with the same quantum number $J^P$. We take $s_0^A  = 33.0(1)$, $s_0^{V_1} = 33.0(1)$, $s_0^{V_2} = 33.0(1)$ and $s_0^{V_{30}} = 33(1)~{\rm GeV}^2$. The selection of the Borel window aims to ensure that the continuum states contribute less than $30\%$ and the contribution from high-twist states should be as small as possible. The determined Borel parameters are ${\cal M}^2_A = 4.0(5)$, ${\cal M}^2_{V_1} = 4.0(5)$, ${\cal M}^2_{V_2} = 16.0(5)$, and ${\cal M}^2_{V_{30}} = 16.0(5)~{\rm GeV}^2$, respectively. We present the $B\to b_1(1235)$ TFFs together with their errors at the large recoil point, {\it i.e.,} $q^2 \to 0~{\rm GeV}^2$, in Table~\ref{Tab:TFFs0}. To make a comparison, the predictions from various approaches are also presented, {\it i.e.,} the LCSR-I~\cite{Yang:2008xw}, the LCSR-II~\cite{Sun:2011ssd}, the pQCD~\cite{Li:2009tx}, and the CLFQM~\cite{Cheng:2003sm}, respectively. Table~\ref{Tab:twist} shows how different twist LCDAs affect the $B\to b_1(1235)$ TFFs at the large recoil point $q^2=0$.

\begin{table}[htb]
\caption{The masses of low-lying $B$ resonances, coefficients $\alpha_{1,2}$, and $\Delta$ for the TFFs $A(q^2), V_1(q^2), V_2(q^2), V_0(q^2)$, in which all the input parameters are set to be their central values.} \label{Tab:SSE}
\begin{tabular}{c c c c c}
\hline
~~~~~~~~~~~~        & ~~~$A(q^2)$~~~     & ~~~$V_1(q^2)$~~~   & ~~~$V_2(q^2)$~~~   & ~~~$V_0(q^2)$~~~  \\   \hline
$m_{R,i}$ & 5.324 & 5.726 & 5.726 & 5.279 \\
$\alpha_1$    & $+0.672$   & $-0.529$   & $+0.019$   & $-4.493$  \\
$\alpha_2$    & $+1.684$   & $+1.185$  & $+5.513$   & $-11.685$  \\
$\Delta$      & $-0.037\%$  & $-0.012\%$  & $-0.049\%$  & $-0.009\%$   \\
\hline
\end{tabular}
\end{table}

\begin{figure}[htb]
\begin{center}
\includegraphics[width=0.40\textwidth]{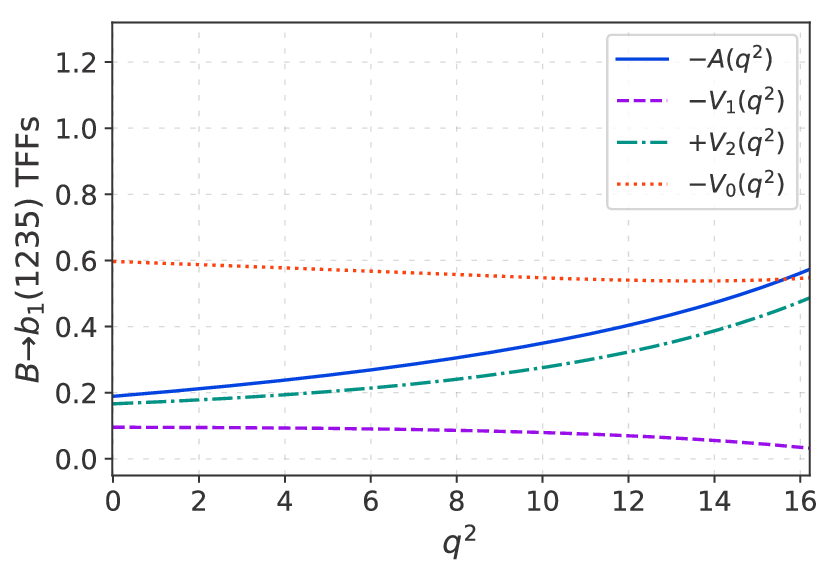}
\end{center}
\caption{The $B\to b_1(1235)$ TFFs $A(q^2)$, $V_1(q^2)$, $V_2(q^2)$ and $V_0(q^2)$ using the central values of all the input parameters.}
\label{fig:TFFs}
\end{figure}

Theoretically, the LCSR approach for $B\to b_1(1235)$ TFFs is reliable in low and intermediate $q^2$ regions, which can be extrapolated to all physically allowable regions $m_\ell^2 \leq q^2 \leq (m_B - m_{b_1})^2 \approx 16.40~{\rm GeV}^2$ via a proper way. Here we adopt the simplified series expansion approach~\cite{Bourrely:2008za, Bharucha:2010im} to do the extrapolation, which takes the following form:
\begin{eqnarray}
F_i(q^2) = \frac1{P_i(q^2)}\sum\limits_k \alpha_k [z(q^2)-z(0)]^k.
\end{eqnarray}
Here the function $z(t)$ is defined as
\begin{eqnarray}
z( t) = \frac{{\sqrt {{t_ + } - t}  - \sqrt {{t_ + } - {t_0}} }}{{\sqrt {{t_ + } - t}  + \sqrt {{t_ + } - {t_0}} }},
\end{eqnarray}
where ${t_\pm} = {({m_B} \pm{m_{{b_1}}})^2} and {t_0} = {t_ + }(1 - \sqrt {1 - {t_ {-}}/{t_{ + }}} )$. $F_i(q^2)$ are the TFFs $A(q^2)$ and $V_{0,1,2}(q^2)$, respectively. In this approach, the simple pole $P_i(q^2)=(1-q^2 / m^2_{R,i})$, which accounts for the low-lying resonance, is adopted. Here, the masses of low-lying $B$ resonances are mainly determined by the $J^P$ states. Followed by Ref.~\cite{Momeni:2020zrb} and PDG values~\cite{Zyla:2020zbs}, we listed the $m_{R,i}$ in Table~\ref{Tab:SSE}. The input parameters are fitted by requiring $\Delta<1\%$, where the parameter $\Delta$ is usually introduced to measure the quality of fit, which is defined as
\begin{align}
\Delta  = \frac{\sum\nolimits_t|F_i(t)-F_i^{\rm fit}(t)|}{ \sum\nolimits_t |F_i(t)|}\times  100,
\end{align}
where $t \in [0,1/40, \cdots ,40/40]\times  8.20~{\rm GeV}^2$. The fitting parameters $\alpha_i$ for every TFF and the quality of fit $\Delta$ are also listed in Table~\ref{Tab:SSE}. It shows that the $\Delta$ of $B\to b_1(1235)$ TFFs are less than $0.050\%$, indicating high quality of fit. We present the extrapolated  $B\to b_1(1235)$ TFFs $A(q^2)$, $V_1(q^2)$, $V_2(q^2)$, and $V_0(q^2)$in Fig.~\ref{fig:TFFs}, in which the TFFs $A(q^2)$, $V_1(q^2)$, and $V_0(q^2)$ are added with a minus sign for convenience.

\begin{figure*}[t]
\includegraphics[width=0.45\textwidth]{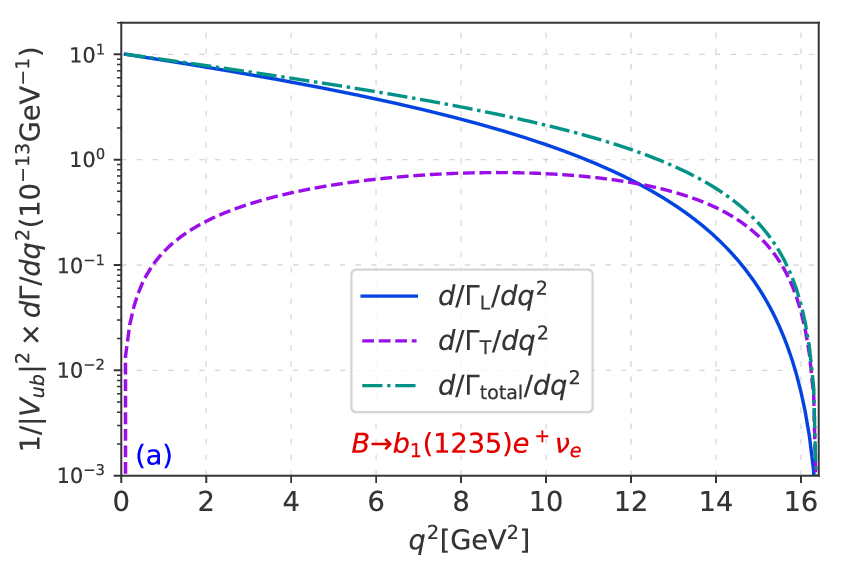}\includegraphics[width=0.45\textwidth]{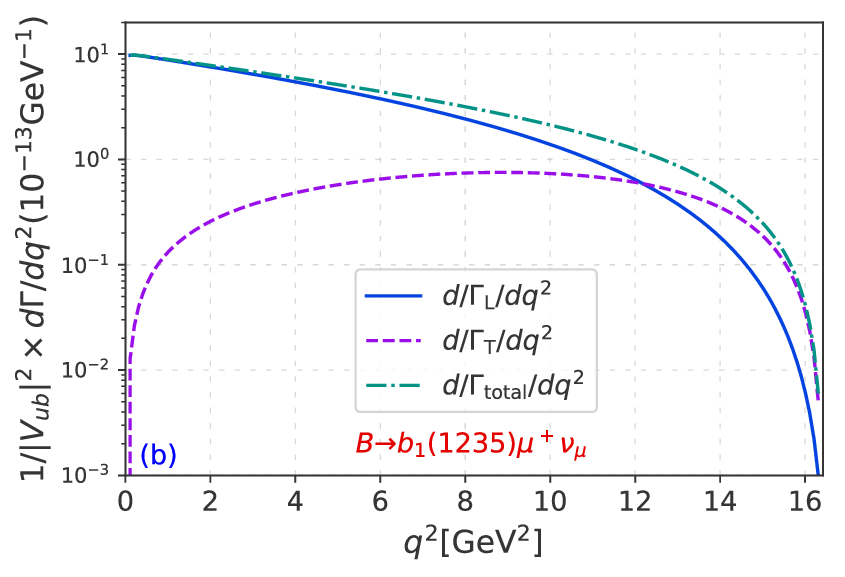}
\includegraphics[width=0.45\textwidth]{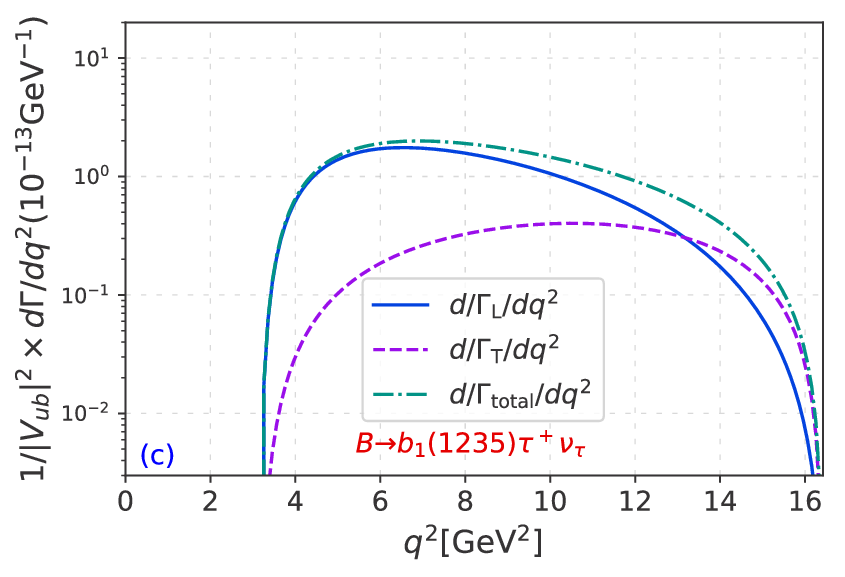}
\caption{The $B \to b_1(1235) \ell^+ \nu_\ell$ decay width versus $q^2$ by using the TFFs derived from the LCSR approach, where $\ell= (e,\mu,\tau)$, respectively. Among them, figure (a) is the $B \to b_1(1235) e^+ \nu_e$ decay channel, figure (b) is the $B \to b_1(1235) \mu^+ \nu_\mu$ decay channel, and figure (c) is the $B \to b_1(1235) \tau^+ \nu_\tau$ decay channel, respectively}
\label{Fig:dG}
\end{figure*}

\begin{table*}[htb]
	\begin{center}
		\caption{The branching fractions of $B \to b_1(1235)\ell^+\nu_\ell$ with errors (in units $10^{-4}$).}
		\label{tab:bf}
		\begin{tabular}{lcc}
			\hline
			~~~~~~~~~~~~~~~~~& ${\cal B}(\bar B^0\to b_1(1235)^+e^- \bar\nu_e) $~~~~~~~~ & ${\cal B}(B^+\to b_1(1235)^0e^+\nu_e)$\\
			\hline
			This work    & $2.179^{+0.553}_{-0.422}$ & $2.353^{+0.597}_{-0.456}$    \\
			LCSR~\cite{Yang:2008xw}  & $1.93^{+0.84}_{-0.68}$ & $2.07^{+0.90}_{-0.73}$ \\
			pQCD~\cite{Li:2009tx}      & $2.88^{+1.51}_{-1.22}$  & ... \\
			CLFQM~\cite{Kang:2018jzg}   &  ...            & $0.77\pm0.17$ \\
			\hline
			~~~~~~~~~& ${\cal B}(B^0\to b_1(1235)^-\mu^+\nu_\mu)$~~~~   & ${\cal B}(B^+\to b_1(1235)^0\mu^+\nu_\mu)$\\
			\hline
			This work    &   $2.166^{+0.544}_{-0.415}$ & $2.339^{+0.587}_{-0.448}$  \\
			pQCD~\cite{Li:2009tx}      & $1.26^{+0.66}_{-0.54}$  & ... \\
			\hline
		\end{tabular}
	\end{center}
\end{table*}

To remove the uncertainty from the CKM matrix element $|V_{ub}|$, according to Eqs.(\ref{Eq:dgamma1}) and (\ref{Eq:dgamma2}), the longitudinal and transverse differential decay widths $d\Gamma_{\rm L,T}$ and total width $d\Gamma_{\rm total} = d\Gamma_{\rm L} + d\Gamma_{\rm T}$ of $B \to b_1(1235)\ell^+ \nu_\ell$ need to be divided by $|V_{ub}|^2$. We present their behavior as a function of $q^2$ in Fig. \ref{Fig:dG}. Figures \ref{Fig:dG}(a)--\ref{Fig:dG}(c) show the differential widths for the $e^+ \nu_{e^+}$, $\mu^+ \nu_{\mu^+}$, and $\tau^+ \nu_{\tau^+}$ channels, respectively. In all those channels, the transverse contributions $d\Gamma_{\rm T}/dq^2$ are relatively small compared to the longitudinal contributions $d\Gamma_{\rm L}/dq^2$ in the low and intermediate $q^2$ regions. Thus, total differential widths mainly come from the longitudinal components.

Finally, by using the lifetimes of $B^0, B^+$ mesons ${\tau_{B^0}} = (1.517 \pm 0.004)$ and $\tau_{B^+} = (1.638 \pm 0.004)~\rm ps$ issued by the PDG~\cite{Zyla:2020zbs}, we predict the branching fractions for the two different semileptonic decay channels $B^0 \to b_1(1235)^- \ell^+ \nu_\ell$ and $B^+ \to b_1(1235)^0 \ell^+ \nu_\ell$, which are listed in Table~\ref{tab:bf}. Meanwhile, other theoretical predictions such as pQCD~\cite{Li:2009tx}, LCSR~\cite{Yang:2008xw}, and CLFQM~\cite{Cheng:2003sm} are also given as a comparison. Our results align well with the previous LCSR predictions~\cite{Yang:2008xw} within reasonable error ranges.

\section{Summary}\label{Sec_IV}

In the present paper, we have calculated the $b_1(1235)$-meson $\xi$ moments of LCDA $\langle\xi_{2;b_1}^{n;\|}\rangle$ by using the BFTSR approach up to dimension-six condensates. The first two nonzero moments of LCDA have been given in Eq. (\ref{Eq:xi}). Figure \ref{fig:DA1} shows that $\phi_{2;b_1}^\|$ tends to be an antisymmetric form. Figure \ref{fig:DA2} shows the scale-running behavior of $\phi_{2;b_1}^\|(x)$. Figure \ref{fig:DA3} shows how the LCDA $\phi_{2;b_1}^\|$ at $1$ GeV changes when more Gegenbauer terms are included, {\it e.g.,} the Gegenbauer moments are kept up to $n=1$ and $n=3$, respectively. Moreover, by using the derived leading-twist LCDA, we have calculated the $B\to b_1(1235)$ TFFs $A(q^2)$ and $V_{0,1,2}(q^2)$ by using the LCSR approach up to twist-four accuracy. At largest recoil point, we obtain $A(0) = -0.133^{+0.026}_{-0.033}$, $V_1(0) = -0.101_{-0.016}^{+0.013}$, and $V_2(0) = +0.344_{ - 0.049}^{+0.054}$. Furthermore, by extrapolating the TFFs to its physical allowable $q^2$ range, we present the $|V_{ub}|$-free differential decay widthes of the semileptonic decays $B \to b_1(1235)\ell^+\nu_\ell$ with $\ell = (e,\mu, \tau)$ in Fig.\ref{Fig:dG}. The branching fractions for $B^{0(+)} \to b_1(1235)^{-(0)}\ell^+\nu_\ell$ have been given in Table~\ref{tab:bf}. It is hoped that the decays $B\to b_1(1235)\ell^+\nu_\ell$ can be observed in near future, which inversely could provide a (potential) helpful test for the QCD sum rules approach.

\hspace{1cm}

{\bf Acknowledgments:} H.-B. F. and T. Z. would like to thank the Institute of High Energy Physics of Chinese Academy of Sciences for their warm and kind hospitality. This work was supported in part by the National Natural Science Foundation of China under Grants No. 12175025, No. 12265010, No. 12265009, and No. 12347101, the Project of Guizhou Provincial Department of Science and Technology under Grants No. ZK[2025]MS219 and No. ZK[2023]024, the Graduate Research and Innovation Foundation of Chongqing, China under Grants No. CYB23011 and No. ydstd1912, and the Fundamental Research Funds for the Central Universities under Grant No. 2024IAIS-ZD009.

\end{document}